\documentclass[aps,prb,twocolumn,superscriptaddress,showpacs,floatfix]{revtex4}
\usepackage[colorlinks=true,urlcolor=blue,citecolor=blue,linkcolor=blue]{hyperref}
\usepackage{graphicx}
\usepackage{latexsym}
\usepackage{stmaryrd}
\usepackage{amsmath}
\usepackage{amssymb}
\usepackage{epstopdf}
\usepackage{slashed}
\usepackage{mathrsfs}
\usepackage{bm}
\usepackage{verbatim}
\usepackage{hyperref}
\usepackage{dsfont}

\begin{document}
\setcitestyle{numbers,square}
\makeatletter
\newcommand{\rmnum}[1]{\romannumeral #1}
\newcommand{\Rmnum}[1]{\expandafter\@slowromancap\romannumeral #1@}

\makeatother

\title{Quantum Hall Effect in a 3D Cross-Dimensional Quantum Anomalous Hall System}
\author{Liang-Liang Cheng}
\affiliation{Interdisciplinary Center for Theoretical Physics and Information Sciences (ICTPIS) and State Key Laboratory of Surface Physics, Fudan University, Shanghai 200433, China}
\affiliation{Department of Physics, Fudan University, Shanghai 200433, China}
\author{Ao-Long Li}
\affiliation{Interdisciplinary Center for Theoretical Physics and Information Sciences (ICTPIS) and State Key Laboratory of Surface Physics, Fudan University, Shanghai 200433, China}
\affiliation{Department of Physics, Fudan University, Shanghai 200433, China}
\author{Zhi-Qiang Zhang}\email{zhangzhiqiangphy@163.com}
\affiliation{Interdisciplinary Center for Theoretical Physics and Information Sciences (ICTPIS) and State Key Laboratory of Surface Physics, Fudan University, Shanghai 200433, China}
\affiliation{School of Physical Science and Technology, Soochow University, Suzhou, 215006, China}
\author{Hua Jiang}\email{jianghuaphy@fudan.edu.cn}
\affiliation{Interdisciplinary Center for Theoretical Physics and Information Sciences (ICTPIS) and State Key Laboratory of Surface Physics, Fudan University, Shanghai 200433, China}
\date{\today}

\begin{abstract}

We systematically investigate cross-dimensional transport and the resulting quantum Hall effect in 3D quantum anomalous Hall effect systems, which are constructed by coupling two Weyl-semimetal Hamiltonians. Our study unveils that the system hosts highly anisotropic surface states and chiral modes, leading to rich quantum Hall responses. Crucially, we clarify how magnetic fields applied along three high-symmetry directions dictate the evolution of the Chern numbers, chiral modes, and Hall responses. We discover that chiral modes along different directions can originate from distinct mechanisms, giving rise to various magnetic modulation characteristics. Furthermore, we unveil the breakdown of the conventional Chern-number-based bulk-boundary correspondence under magnetic fields, where the Chern numbers alone fail to fully characterize the evolution of the chiral states. This anomalous behavior is attributed to the unique momentum-space distributions of surface bands. These insights are further validated numerically through multi-terminal Hall transport simulations. 
Our findings expand the understanding of magnetic-field-driven mechanisms 
in 3D topological systems and provide deeper insights into the interplay between quantized Hall 
responses and Chern numbers in 3D systems.
\end{abstract}
\maketitle

\section{Introduction}

The quantum Hall effect provides one of the most fundamental examples of topological phases of matter, in which the Hall conductance is quantized and protected by a topological invariant \cite{vonKlitzing1980FineStructureConstant,
TKNN1982,Niu1985HallTopologicalInvariant}. In two-dimensional systems, the integer quantum Hall effect is characterized by a nonzero Chern number, and the bulk-boundary correspondence guarantees the existence of chiral boudary states at the sample boundary \cite{TKNN1982,Halperin1982EdgeStates,Hatsugai1993ChernEdge}. Its zero-magnetic-field counterpart, the quantum anomalous Hall effect, further demonstrates that quantized Hall transport can arise from intrinsic band topology, spin-orbit coupling, and time-reversal-symmetry breaking, without the need for Landau-level formation \cite{Haldane1988ChernInsulator,Yu2010QAH,
Chang2013QAH,Chang2015DissipationlessEdge}. These developments have established Chern-number topology as a central concepts for understanding dissipationless chiral transport in two-dimensional systems.

Extending quantum Hall physics to three dimensions has attracted 
considerable interest, because three-dimensional topological systems 
may host boundary states with richer spatial structures than their 
two-dimensional counterparts
\cite{Halperin1987PossibleStates,
Kohmoto1992Diophantine,Schindler2018HigherOrder,
Tang2019ThreeDimensionalQHE,Zhao2024ThreeDQAH,arXiv.2610.27802}. 
In particular, Weyl semimetals provide a novel platform for the 
realization of three-dimensional quantum Hall effects
\cite{PhysRevLett.119.136806,
PhysRevLett.125.036602,
Chang2021ThreeDQHE,
Ma2021MagnetothermoelectricQHE,
Xiong2022GenericMultiWeylQHE}. 
A Weyl semimetal hosts pairs of Weyl nodes with opposite chirality in 
the bulk Brillouin zone
\cite{Wan2011FermiArc,Burkov2011WSM,
Weng2015TaAs,Armitage2018WeylDiracReview,Inoue2016FermiArcConnectivity}, and these Weyl nodes are connected by 
Fermi-arc surface states. Unlike closed Fermi surfaces in ordinary 
metals, Fermi arcs terminate at the projected Weyl nodes and are 
intrinsically tied to the bulk topology \cite{Wan2011FermiArc,Xu2015WeylDiscovery,
Lv2015TaAs,Armitage2018WeylDiracReview,
Okugawa2014FermiArcDispersion}. 
When the chiral modes on different surfaces are connected through the 
bulk or through boundary coupling, they can form closed trajectories 
in momentum space, giving rise to effective three-dimensional quantum 
Hall responses
\cite{Potter2014FermiArcOrbits,
Zhang2019WeylOrbitQHE,
PhysRevLett.119.136806,
PhysRevLett.125.036602,
Moll2016FermiArcChiralityTransfer,Zhang2017EvolutionWeylOrbitQHE}.

Recently, a three-dimensional quantum anomalous Hall effect has been proposed in systems constructed by Weyl semimetals \cite{Zhang2025ThreeDimensionalQAH}. In such systems, the interplay between Fermi arc surface states and intersurface coupling can generate cross-Dimensional chiral transport. The relevant boundary modes are not confined to a single edge of a two-dimensional plane, but may be distributed over surfaces, hinges, or surface intersections \cite{Schindler2018HigherOrder,Chen2021HingeStates,
Zhang2025ThreeDimensionalQAH,SciAdv.12.eaeb4171}. In particular, the Chern contributions from two opposite surfaces can sum to zero, corresponding to a zero total Chern number when the system is viewed in an effective two-dimensional description, while chiral boundary modes remain present. Moreover, since an individual facet can carry a half-quantized Chern contribution, the conventional two-dimensional boudary state counting based on the difference of Chern numbers cannot in general be applied independently to adjacent facets \cite{Zhang2025ThreeDimensionalQAH}. Since the relevant conducting channels may originate from Fermi-arc-derived surface states and from coupling between different facets, an applied magnetic field can affect not only the Chern numbers of individual surfaces, but also the relative momentum-space alignment of surface bands. It is therefore necessary to clarify whether the magnetic-field-induced transport behavior can still be understood within the conventional Chern-number framework, or whether additional mechanisms are required.

\begin{figure}[h]
  \centering
  \includegraphics[width=\columnwidth]{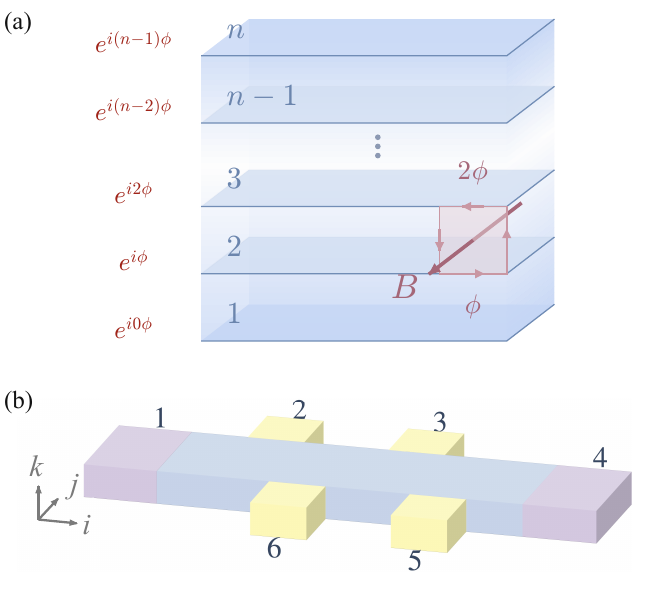}
    \caption{(a) Schematic illustration of the Peierls phases induced by a magnetic field. The phase varies linearly between successive layers, \(\phi_n=(n-1)\phi\). (b) Schematic illustration of the six-terminal Hall bar setup.}
    \label{F1}
\end{figure}

In this work, we investigate the magnetic-field effects on a three-dimensional quantum anomalous Hall effect (3D QAHE) system characterized by cross-dimensional transport, which is constructed by coupling two Weyl semimetal Hamiltonians. We first revisit its surface band structures and the corresponding Chern numbers, providing a more detailed characterization of the surface states. Within the conventional surface classification, the $y$-$z$ surfaces possess nontrivial half-integer Chern numbers, whereas the $x$-$y$ and $x$-$z$ surfaces have vanishing Chern numbers. Importantly, the latter surfaces exhibit a characteristic momentum-surface locking, where surface states with opposite momenta are localized on different surfaces in real space. These features provide a distinctive mechanism for chiral boundary transport that transcends the conventional Chern number description.

We then study the evolution of the chiral boundary states under magnetic fields applied along the three principal directions. For magnetic fields along the $y$ and $z$ directions as illustrate in Fig.~\ref{F1}(a), the magnetic field reconstructs the chiral channels within the framework of conventional Chern number descriptions. However, for chiral modes occurring along directions with zero Chern numbers, the standard conventional Chern-number description can fail even under a magnetic field. Specifically, the chiral modes can evolve spatially without a corresponding change in the Chern number. Instead, the Aharonov-Bohm effect plays a more fundamental role \cite{Aharonov1959Potentials}. It induces a phase shift that displaces the isolated surface bands in momentum space. These shifting surface bands reconfigure the chiral modes across different surfaces, thereby modulating the Hall response even when the Chern numbers remain preserved. These insights are further validated through multi-terminal Hall transport simulations.

The rest of this paper is organized as follows: In Sec. \ref{section2}, we give the details of the model and methods. In Sec. \ref{section3}, we demonstrate the Chern number, the surface states and the chiral modes without magnetic fields. The corresponding formation mechanisms of these chiral modes are clarified accordingly. The effects of magnetic fields and details of the QHE are shown in Sec. \ref{section4}. In Sec. \ref{section5}, the corresponding Hall responses are given. Finally, a brief summary is presented in Sec. \ref{section6}.

\section{Model and method}\label{section2}

\subsection{Model}
Following the previous study, the Hamiltonian of 3D QAHE \cite{Zhang2025ThreeDimensionalQAH} is given by
\begin{equation}
\mathcal{H}
=
\mathcal{H}_w \sigma_z
+
\sum_{n}
\left[
i t_s
\left(
\sigma_y c_n^\dagger c_{n+\delta_y}
+
\sigma_x c_n^\dagger c_{n+\delta_z}
\right)
+
\mathrm{h.c.}
\right],
\label{10}
\end{equation}
where $\sigma_{x,y,z}$ are the Pauli matrices, $c_n^\dagger$ and $c_n$ are the electron creation and annihilation operators, respectively, and $\delta_\alpha$ $(\alpha=x,y,z)$ denotes the nearest-neighbor lattice displacement along the $\alpha$ direction. The Weyl-semimetal Hamiltonian $\mathcal{H}_w$ reads
\begin{equation}
\mathcal{H}_w
=
\sum_n T_0 c_n^\dagger c_n
+
\sum_{\alpha=x,y,z}
T_\alpha c_n^\dagger c_{n+\delta_\alpha}
+
\mathrm{h.c.},
\end{equation}
with
$T_0 = -4m\tau_z - \gamma$, $T_{x/y} = m\tau_z - iA\tau_{x/y}$, $T_z = m_z \tau_z$, where $\tau_{x,y,z}$ are Pauli matrices acting in spin space. The first term in the Hamiltonian describes a Weyl semimetal with two Weyl points, while the second term accounts for the coupling between the two Weyl semimetals and the Rashba spin-orbit interaction that opens a bulk energy gap $\Delta E =2t_s$. Without loss of generality, the system parameters are fixed at
$A=\gamma=t_s=0.5t$, $m=m_z=t$.

In the presence of an external magnetic field, we incorporate only its orbital effect through the Peierls substitution \cite{Peierls1933}. Specifically, each nearest-neighbor hopping process acquires a phase factor,
\begin{equation}
T_{\alpha}
\rightarrow
T_{\alpha} e^{i\theta_{n,\alpha}},
\qquad
\theta_{n,\alpha}
=
\frac{2\pi}{\Phi_0}
\int_{\mathbf r_n}^{\mathbf r_n+\boldsymbol{\delta}_{\alpha}}
\mathbf A(\mathbf r)\cdot d\mathbf l ,
\end{equation}
where $\mathbf A$ is the vector potential satisfying $\mathbf B=\nabla\times\mathbf A$, $\Phi_0=h/e$ is the magnetic flux quantum, and $\theta_{n,\alpha}$ denotes the Peierls phase accumulated when an electron hops from site $n$ to its nearest-neighbor site $n+\delta_\alpha$. Here the onsite term remains unchanged, while all nearest-neighbor hopping terms acquire the corresponding Peierls phases. As plotted in Fig.~\ref{F1}, the gauge-invariant phase accumulated around one unit cell is $\phi = 2\pi\Phi/\Phi_0$, where $\Phi$ is the magnetic flux threading the plaquette.

\begin{figure}[h]
  \centering
    \includegraphics[width=\columnwidth]{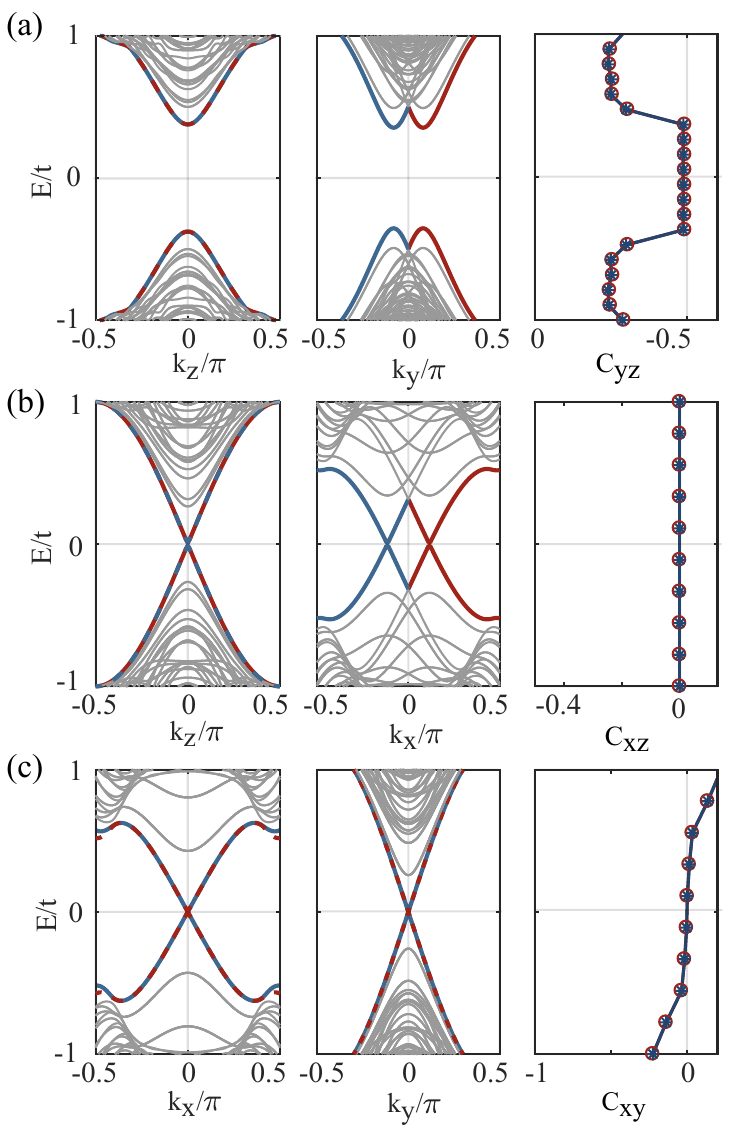}
    \caption{Isolated surface bands and Chern numbers. The red and blue curves mark the surface states closest to zero energy on opposite surfaces. The red and blue lines in the rightmost panel give the corresponding surface Chern numbers.
(a) $y$--$z$ surface; the sample sizes are  $L_x=18$ and $L_y(L_z)=18$ for the spectrum along $z(y)$ direction, $L_y=L_z=18$ and $L_x=6$ for the Chern-number calculation.
(b) $x$--$z$ surface; the sample sizes are $L_y=18$ and $L_x(L_z)=18$ for the spectrum along $z(x)$ direction, $L_x=L_z=18$ and $L_y=6$ for the Chern-number calculation.
(c) $x$--$y$ surface; the sample sizes are $L_z=18$ and $L_y(L_x)=18$ for the spectrum along $x(y)$ direction, $L_x=L_y=18$ and $L_z=6$ for the Chern-number calculation.}
    \label{F2}
\end{figure}

Accordingly, the Hamiltonian under the magnetic field is written as
\begin{equation}
\begin{aligned}
\mathcal{H}(\mathbf B)
&=
\mathcal{H}_w(\mathbf B)\sigma_z\ +
\sum_n
[
i t_s
(
\sigma_y e^{i\theta_{n,y}}
c_n^\dagger c_{n+\delta_y}
\\
&
+
\sigma_x e^{i\theta_{n,z}}
c_n^\dagger c_{n+\delta_z}
)
+
\mathrm{h.c.}
],
\end{aligned}
\end{equation}
with
\begin{equation}
\mathcal{H}_w(\mathbf B)
=
\sum_n T_0 c_n^\dagger c_n
+
\sum_{\alpha=x,y,z}
\left[
T_\alpha e^{i\theta_{n,\alpha}}
c_n^\dagger c_{n+\delta_\alpha}
+
\mathrm{h.c.}
\right].
\end{equation}
The direction of the magnetic field is manipulated by adjusting $\theta_{n,\alpha}$ accordingly.

\subsection{Method}
To characterize the topological nature of the magnetic-field-driven phases, we employ the real-space Chern-number formalism \cite{Bianco2011Mapping,Prodan2011Disordered}. For a given Fermi energy $E_F$ located inside the bulk gap, the occupied-state projection operator is defined as
\begin{equation}
P
=
\sum_{E_m<E_F}
|\psi_m\rangle\langle\psi_m|,
\qquad
Q=1-P,
\end{equation}
where $|\psi_m\rangle$ and $E_m$ are the eigenstates and eigenenergies of $\mathcal H(\mathbf B)$, respectively. The local Chern marker on the plane spanned by the $i$ and $j$ directions [see Fig.~\ref{F1}] is written as
\begin{equation}
C_{ij}(\mathbf r)
=
-4\pi
\,\mathrm{Im}
\sum_{\chi}
\langle \mathbf r,\chi|
P\hat r_i Q\hat r_j P
|\mathbf r,\chi\rangle ,
\label{eq:local_chern_marker}
\end{equation}
where $\hat r_i$ is the position operator along the $i$ direction and $\chi$ denotes the internal degrees of freedom, including the spin and orbit indices. For the three-dimensional system, the Chern vector is defined by
\begin{equation}
\begin{aligned}
C_{ij}
=
\frac{1}{N_{i} N_{j}}
\sum_{\mathbf r\in all}
C_{ij}(\mathbf r),
\label{eq:chern_vector}
\end{aligned}
\end{equation}
with $k,i,j\in \{x,y,z\}$, $N_{i}$, $N_{j}$ and  $N_k$ are sample sizes along $i$, $j$, and $k$ directions, respectively. The surface Chern number for each surface is defined by counting local Chern numbers of the surfaces. For example, for the front surfaces in Fig.~\ref{F1}(a), one counts half of the sample with
\begin{equation}
\begin{aligned}
C_{ij;top}
=
\frac{1}{N_{i} N_{j}}
\sum_{\mathbf r_i,r_j\in all;r_k>N_k/2 }
C_{ij}(\mathbf r).
\label{eq:chern_vector}
\end{aligned}
\end{equation}
A similar formula is available for the opposite surface, except for $r_k<N_k/2$. We emphasize that the surface Chern numbers already include the bulk Chern numbers.

The transport properties are calculated by using the nonequilibrium Green's function method~\cite{Datta1995Electronic,Caroli1971TunnelingCurrent,LI20241221}. The central region described by $\mathcal H_C(\mathbf B)$ is connected to semi-infinite leads, labeled by $m$. For the six terminal setup in Fig.~\ref{F1}(b), the leads are marked from $m=1$ to $m=6$. The retarded Green's function of the central region satisfies
\begin{equation}
G^r(E)
=
\left[
E+i\eta
-
\mathcal H_C(\mathbf B)
-
\sum_{m\in[1,6]}\Sigma_{m}^r(E)
\right]^{-1},
\label{eq:gr}
\end{equation}
where $\eta\rightarrow 0^+$, and $\Sigma_{m}^r(E)$ is the retarded self-energy of the $m$-th lead. The corresponding linewidth function is defined as
\begin{equation}
\begin{aligned}
\Gamma_{m}(E)
=
i
\{
\Sigma_{m}^r(E)
-
[\Sigma_{m}^r(E)]^\dagger
\}.
\end{aligned}
\end{equation}
The transmission coefficient between $m$ and $n$ terminals is given by the Landauer--B\"uttiker formula \cite{Landauer1957Spatial,Buttiker1986FourTerminal,
Datta1995Electronic,Caroli1971TunnelingCurrent}
\begin{equation}
T_{\mathrm{mn}}(E)
=
\mathrm{Tr}
\left[
\Gamma_m(E)
G^r(E)
\Gamma_n(E)
G^a(E)
\right],
\label{eq:transmission}
\end{equation}
with $G^a(E)=[G^r(E)]^\dagger$.

To further analyze the current and voltage distributions, one should solve the Kirchhoff equation \cite{Buttiker1986FourTerminal,PhysRevB.38.9375}
\begin{equation}
I_m
=
\frac{e^2}{h}
\sum_{n\ne m}
\left(
T_{nm}V_m
-
T_{mn}V_n
\right).
\label{eq:lb_kirchhoff}
\end{equation}
$V_m$ is the bias of terminal $m$ and $T_{mn}$ denotes the transmission probability from terminal $n$ to terminal $m$.
After solving Eq.~\eqref{eq:lb_kirchhoff}, the Hall resistance can be obtained from
\begin{equation}
R_{ij}
=
\frac{V_2-V_6}{I}.
\end{equation}
A current $I$ is injected into the sample through terminal $1$ and extracted through terminal $4$. $V_2$ and $V_6$ are the voltages measured at terminals $2$ and $6$, respectively [see Fig.~\ref{F1}(b)].
The longitudinal resistance can be obtained from
\begin{equation}
R_{ii}
=
\frac{V_2-V_3}{I}.
\end{equation}

\section{Band properties without magnetic fields}\label{section3}

\begin{figure}[h]
  \centering
    \includegraphics[width=\columnwidth]{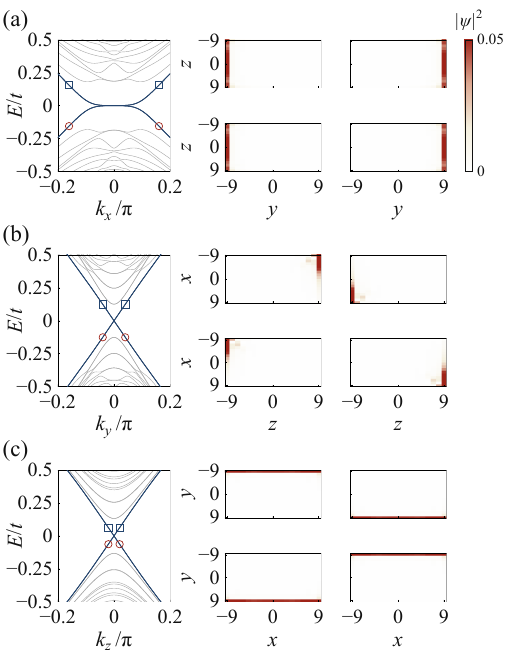}
    \caption{(a) Energy dispersion $E(k_x)$; the sample sizes are $L_y=L_z=18$.
(b) Energy dispersion $E(k_y)$; the sample sizes are $L_x=L_z=18$.
(c) Energy dispersion $E(k_z)$; the sample sizes are $L_x=L_y=18$.
The blue solid lines indicate chiral states. The blue squares and red circles mark the momenta whose corresponding wave-function distributions are shown to the right of each band structure, in the same order.}
    \label{F3}
\end{figure}

% To characterize the topological properties of the surface and bulk states in this model,
% while avoiding interference from irrelevant surfaces, periodic boundary conditions are imposed only along the lateral directions when analyzing the properties of an individual surface \cite{Xiao2021TopologicalQuantumMaterials,Yoo2021FiniteSizeCorrection}.

To characterize the effects of magnetic fields, it is instructive to first clarify the properties of the surface and bulk states in the absence of a magnetic field. The essential zero-field boundary-state properties of this 3D QAHE system were reported in our previous Short Communication [32], where their overall configurations and transport signatures were briefly summarized. Here, we provide a detailed analysis of the isolated surface spectra, surface Chern numbers, and the formation mechanisms of the chiral boundary modes, which are essential for understanding their evolution under magnetic fields. The corresponding band structures and Chern numbers for each surface are presented in Fig.~\ref{F2}  \cite{Zhang2025ThreeDimensionalQAH}.
To achieve isolated surface properties, open boundary conditions are applied exclusively along the surfaces' stacking direction.
For the $y$-$z$ surface in Fig.~\ref{F2}(a), a pronounced surface band gap is observed, yielding a half-integer Chern number of $C_{yz}=-0.5$ for each gapped surface.
In contrast, the $x$-$z$ and $x$-$y$ surfaces possess gapless surface states [Figs.~\ref{F2}(b) and (c)], where the corresponding Chern numbers vanish. Within an isolated-surface Chern-number description,  these surfaces are topologically trivial. Nevertheless, these two surfaces possess nontrivial roles in the full three-dimensional boundary-state structure  \cite{Zhang2025ThreeDimensionalQAH}.

Despite a vanishing Chern number ($C_{xz}=0$), the $x$-$z$ surface bands display a remarkable feature where the states on opposing surfaces are displaced along opposite $k_x$ directions [see Fig.~\ref{F2}(b)]. Specifically, surface states with $k_x>0$ and $k_x<0$ localize on distinct surfaces in real space, yielding a direct locking between momentum and real-space surface position. The $y$-$z$ surface exhibits a similar momentum-surface locking behavior  \cite{Zhang2025ThreeDimensionalQAH}.

\begin{figure*}[t]
  \centering
  \includegraphics[width=\textwidth]{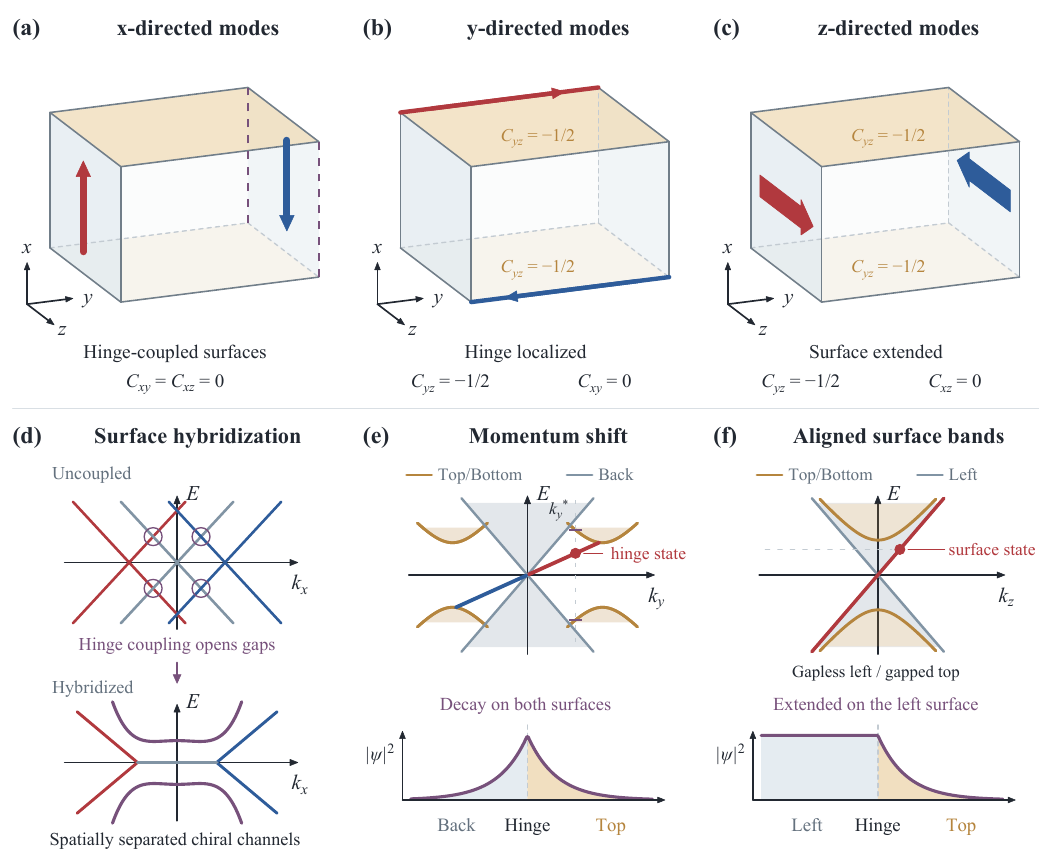}
  \caption{Schematic illustration of chiral boundary states and their formation mechanisms in the absence of magnetic fields.
  (a)--(c) Spatial distributions of chiral surface states along $x$, chiral hinge states along $y$, and chiral side-surface states along $z$, respectively. The purple dashed lines in (a) indicate representative hinges where neighboring side-surface states couple.
  (d) Surface-band dispersions before and after hybridization. Coupling opens gaps near band crossings, leaving spatially separated chiral channels along $x$ despite $C_{xy}=C_{xz}=0$.
  (e) Upper: surface-band dispersions versus $k_y$, with the top and bottom surface bands oppositely shifted by momentum--surface locking. The shaded region marks the common direct gap of the adjacent surfaces. Lower: the selected hinge state decays exponentially into both the back and top surfaces.
  (f) Upper: the gapless side-surface bands and gapped top/bottom surface bands are aligned along $k_z$. Lower: the corresponding chiral state extends over the left surface and decays into the top surface.
  The band dispersions and probability-density profiles are qualitative schematics.}
  \label{F4}
\end{figure*}

To fully characterize the 3D QAHE properties, we next present the chiral state configuration. To achieve this, we adopt a geometry with periodic boundary conditions along one direction and open boundary conditions along the remaining two directions. The corresponding energy spectra are displayed in Fig.~\ref{F3}.

\begin{figure}[h]
  \centering
    \includegraphics[width=\columnwidth]{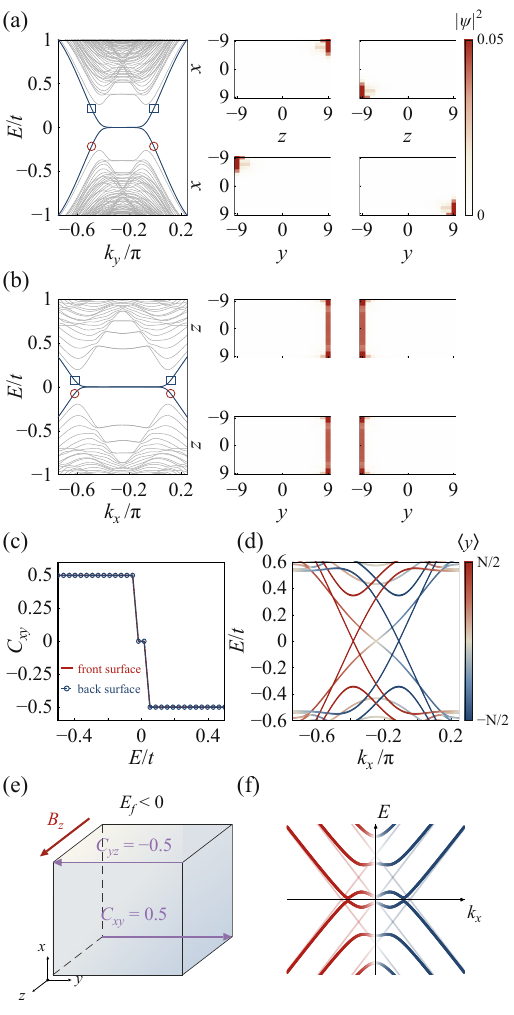}
    \caption{Effects of a magnetic field along the $z$ direction.
(a) Energy dispersion $E(k_y)$; the sample sizes are $L_x=L_z=18$.
(b) Energy dispersion $E(k_x)$; the sample sizes are $L_y=L_z=18$.
(c) Chern number of the $x$--$y$ surfaces; the sample sizes are $L_x=L_y=18$ and $L_z=6$.
(d) Response of the isolated surface bands to the magnetic field.
(e) and (f) Schematic illustrations of how the magnetic field modifies the formation of chiral boundary states propagating along the $y$ and $x$ directions, respectively.
The magnetic phase is $\phi=0.5\pi/17$.}
    \label{F5}
\end{figure}

For nanoribbons oriented along the $z$ and $y$ directions [Figs.~\ref{F3}(b) and (c)], the chiral modes are fully characterized by the corresponding Chern numbers. For the $z$-oriented nanoribbon, the surface Chern number $C_{yz}=-0.5$ and $C_{xz}=0$ give rise to chiral surface transport states. The $x$-$z$ surface exhibits a gapless dispersion at a fixed $k_z$, whereas the $y$-$z$ surface possesses a finite energy gap, as illustrated in Fig.~\ref{F4}(f). The combined effect of these gapless and gapped surface states determines the transport behavior of the system, reminiscent of the chiral boudary states in magnetic topological-insulator thin films \cite{Yu2010QAH,Chang2013QAH,Chang2015DissipationlessEdge}. In contrast, for transport along the $y$ direction, although the relevant surfaces carry the same Chern numbers, the distinct distribution of Dirac cones constrains the chiral modes to the hinges, as shown in Fig.~\ref{F4}(e) \cite{Zhang2025ThreeDimensionalQAH}. These hinge states are associated with adjacent surfaces carrying Chern contributions of $-0.5$ and $0$, respectively. Because the corresponding Chern-number difference is half-integer rather than integer, their origin differs from that of conventional chiral modes at a two-dimensional Chern-number domain wall \cite{Schindler2018HigherOrder,
Langbehn2017SecondOrderTI,
Queiroz2019HingeMode}. These features have been clarified in our previous study  \cite{Zhang2025ThreeDimensionalQAH}.

The chiral modes along the $x$ direction exhibit qualitatively different features. Although all four boundary surfaces carry vanishing Chern numbers [Figs.~\ref{F2}(b) and (c)], stable chiral boudary states still persist, as shown in Fig.~\ref{F3}(a). This observation directly indicates that the Chern-number description alone is insufficient to characterize the chiral boundary states in this system.

\begin{figure}[h]
  \centering
    \includegraphics[width=\columnwidth]{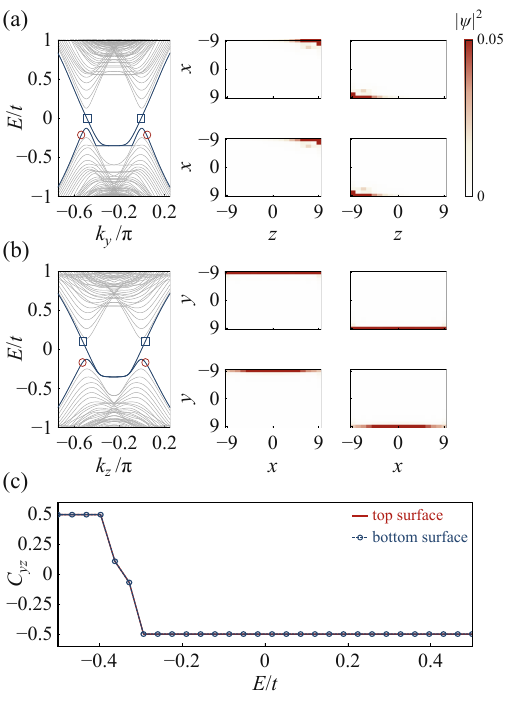}
    \caption{Magnetic field along the $x$ direction.
(a) Energy dispersion $E(k_y)$; the sample sizes are $L_x=L_z=18$.
(b) Energy dispersion $E(k_z)$; the sample sizes are $L_x=L_y=18$.
(c) Chern number of the $y$--$z$ surface; the sample sizes are $L_y=L_z=18$ and $L_x=6$.
The magnetic phase used in the numerical calculations is $\phi=0.5\pi/17$.}
    \label{F6}
\end{figure}

Our previous study demonstrated that such zero-Chern-number chiral boudary states originate from the coupling between different surfaces in a three-dimensional system \cite{Zhang2025ThreeDimensionalQAH}.
When a full transport geometry is considered, open boundary conditions are imposed along both the $y$ and $z$ directions. Under this configuration, the neighboring side surface states in Fig.~\ref{F4}(d) hybridize near the edges.
When the neighboring surface states are coupled at the hinges, the isolated surface spectra in Figs.~\ref{F2}(b) and (c) are reconstructed rather than simply superposed. As schematically illustrated in Fig.~\ref{F4}(d), the band crossings present in the decoupled-surface picture are lifted, and dispersive low-energy bands emerge. The resulting low-energy edge modes mainly inherit their spectral weight from the corresponding side-surface band [$x$-$z$ surface]. This spectral inheritance is fully consistent with the real-space chiral state distribution observed in Fig.~\ref{F3}(a).

\section{Magnetic Field Effects}\label{section4}

Having clarified the band properties without magnetic fields, in the following, we study the magnetic field effect of the above 3D QAHE system. We focus on magnetic fields applied along the three principal crystallographic directions.

\begin{figure}[h]
  \centering
    \includegraphics[width=\columnwidth]{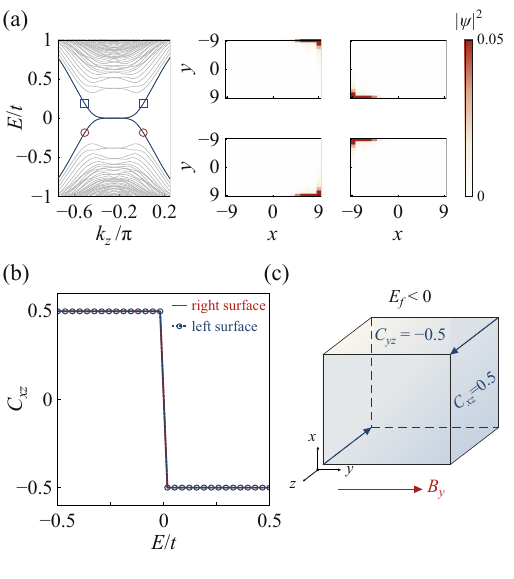}
    \caption{Panels (a), (b) and (c) show the chiral boundary states propagating along the $z$ direction in a $y$-oriented magnetic field, together with the underlying mechanism for their formation. (a) Energy dispersion $E(k_z)$ in the presence of a magnetic field applied along the $y$ direction. (b) Chern number of the $x$-$z$ surface. Panels (c) shows a schematic illustration of how the corresponding magnetic fields modify the formation mechanisms of chiral boundary states propagating along the $z$ direction. In the numerical calculations, the magnetic phase is set to $\phi=0.5\pi/17$ with system sizes $L_x=L_y=18$.}
    \label{F7}
\end{figure}

\subsection{magnetic field along  the $z$ direction}
We first consider the case in which the magnetic field is applied along the $z$ direction, as shown in Fig.~\ref{F5}(e). The $z$-directed magnetic field drives the surface Chern number of the $x$-$y$ surface Dirac cones from $C_{xy} = 0$ to $C_{xy}=\pm0.5$ [see Fig.~\ref{F5}(c)]. 
Owing to the combination of the surface Chern numbers $C_{xy}=\pm0.5$ and $C_{yz}=-0.5$, chiral hinge states along the $y$ direction are observed in Fig.~\ref{F5}(a). These hinge states originate from integer Chern-number differences, which is distinct from the mechanism governing the zero-field ($B=0$) hinge states in Fig.~\ref{F4}(e).

\begin{figure*}[t]
  \centering
    \includegraphics[width=\textwidth]{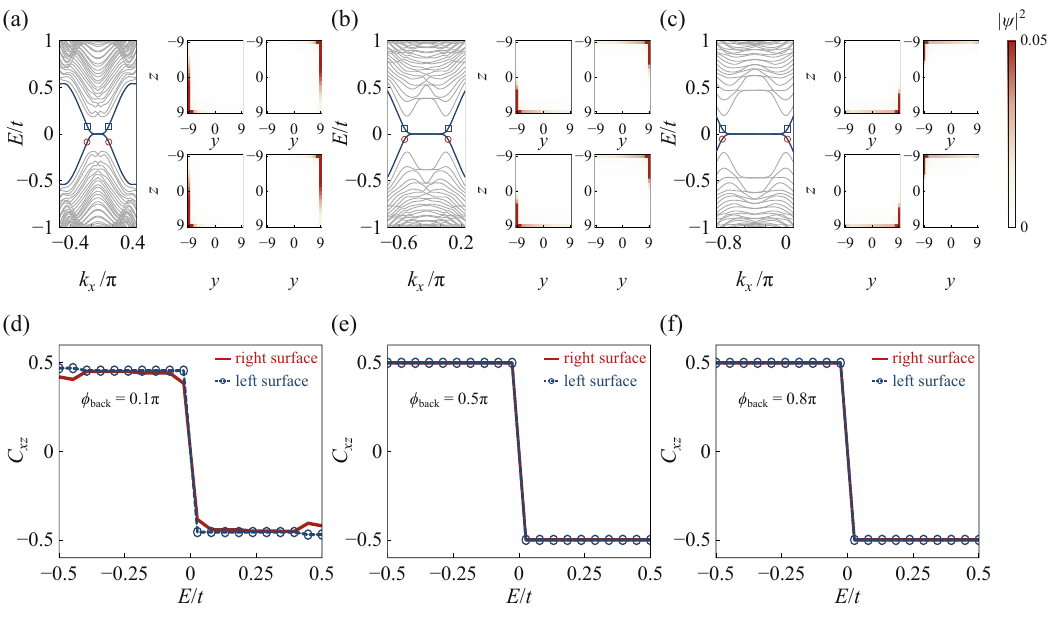}
    \caption{Magnetic field along the $y$ direction. Panels (a), (b), and (c) show the energy dispersion $E(k_x)$ under magnetic fields with different $\phi$: (a)  $\phi=0.1\pi/17$; (b) $\phi=0.5\pi/17$; (c) $\phi=0.8\pi/17$. (d)-(f) Chern numbers of the $x$-$z$ surface  with $\phi=0.1\pi/17$, $\phi=0.5\pi/17$ and $\phi=0.8\pi/17$, respectively. In the numerical calculations, the system sizes are $L_y=L_z=18$.}
    \label{F8}
\end{figure*}

Furthermore, the Aharonov–Bohm phase reconstructs the surface band structure, rendering the conventional Chern-number description incomplete. For the $x$-directed nanoribbon under the same magnetic field [Fig.~\ref{F5}(b)], the emergence of chiral modes stems from the magnetic-flux-induced shift of surface bands along $k_x$. The calculated dispersions of isolated surface states [Fig.~\ref{F5}(d)] reveal that the Dirac cones of the left and right surfaces ($\langle y \rangle=\pm N/2$) are positioned at $k_x=-0.382\pi$ and $k_x=-0.118\pi$, whereas the front and back surface Dirac cones reside at $k_x=-0.25\pi$. This structure is essentially the same as Fig.~\ref{F4}(d), with the primary difference being the magnetic-field-induced shifts in the distance between Dirac cones [see Fig.~\ref{F5}(f)]. Consequently, these shifted cones host the chiral surface transport observed in Fig.~\ref{F5}(b), consistent with the zero-field mechanism.

\subsection{magnetic field along  the $x$ direction}

We next consider the case in which the magnetic field is applied along the $x$ direction. In this configuration, the magnetic field shifts the Dirac cones as well. However, the relative position of the Dirac cones remains unchanged. Similarly to those for B = 0, the localization properties of the chiral boundary states in Figs.~\ref{F6}(a)-(b) are consistent with those in Figs.~\ref{F3}(b)-(c). Specifically, the z-directional (y-directional) chiral modes are located at the surface (hinge) of the sample.

In the zero-field case, the Chern number $C_{yz}$ is fixed at $-0.5$ over the energy range considered, thereby imposing a fixed chirality on the boudary states. By contrast, when the magnetic field is applied along the $x$ direction, $C_{yz}$ varies with the variation of Fermi energy, switching between $0.5$ and $-0.5$ [see Fig.~\ref{F6}(c)]. The energy dependence of $C_{yz}$ drives a chirality reversal for boundary states along the $y$ and $z$ directions. As illustrated in Fig.~\ref{F6}(b), the boundary states at Fermi energies corresponding to $C_{yz}=0.5$ have the opposite chirality to those corresponding to $C_{yz}=-0.5$.

\subsection{magnetic field along  the $y$ direction}

Finally, we study the case in which the magnetic field is applied along the $y$ direction. For the configuration with the magnetic field along the $y$ direction and the nano-ribbon extended along the $z$ direction, the chiral surface states shown in Fig.~\ref{F3}(c) are modified into chiral hinge states shown in Fig.~\ref{F7}(a). This variation also originates from the modification of the Chern number $C_{xz}$. For $E_F<0$, $C_{xz}=0.5$ and $C_{yz}=-0.5$, so the adjacent $x$-$z$ and $y$-$z$ surfaces form an integer Chern-number domain wall, giving rise to the chiral hinge states shown in Fig.~\ref{F7}(c).

For the $x$-directional nano-ribbon in the absence of a magnetic field, chiral modes exist despite the vanishing Chern numbers in both the $x$–$z$ and $x$–$y$ planes. Remarkably, these Chern-number-independent chiral boudnary states depend crucially on the relative positions of the surface bands. Driven by the alignment of the Dirac cones, this unique chiral boudnary state mechanism manifests distinct physical behaviors when a magnetic field is applied along the $y$ direction.

\begin{figure}[h]
  \centering
    \includegraphics[width=\columnwidth]{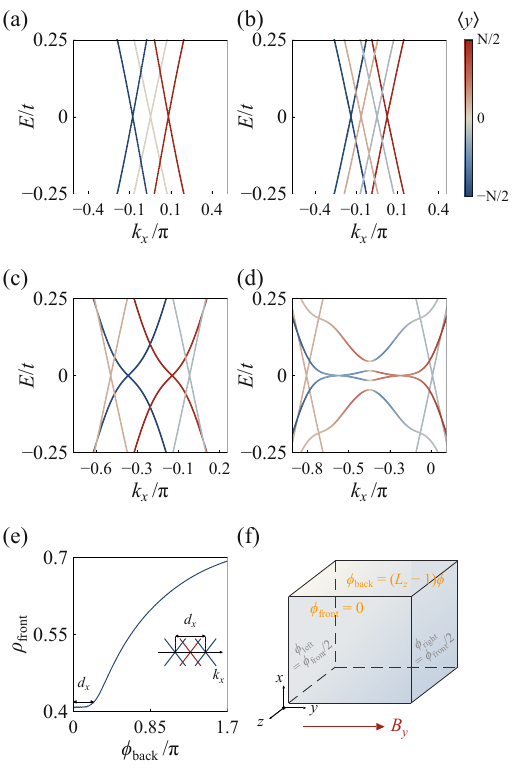}
    \caption{Magnetic-field-induced phenomena in the 3D QAHE without Chern number for transport along the $x$ direction. Panels~(a)--(d) show the response of the uncoupled surface bands to magnetic fields of different strengths: (a) $\phi=0$; (b) $\phi=0.1\pi/17$; (c) $\phi=0.5\pi/17$; (d) $\phi=0.8\pi/17$. (e) Dependence of the transmission density on the front surface on the magnetic-field strength for transport along the $x$ direction under a $y$-directed magnetic field. (f) Schematic of the distinct magnetic-field-induced phases of the surface bands.  In the numerical calculations, the system sizes are $L_y=L_z=18$.}
    \label{F9}
\end{figure}

Specifically, the magnetic field introduces facet-dependent Aharonov-Bohm phases, thereby modulating the relative alignment of the surface bands. For clarity, we plot the evolution of the chiral boudnary states along the $x$ direction under varying magnetic field strengths in Figs.~\ref{F8}(a)–(c). With an increase in $\phi$, the chiral states gradually shift from the $x$-$y$ surfaces to the $x$-$z$ surfaces. Meanwhile, the Chern number $C_{xz}$ remains independent of $\phi$, as shown in Figs.~\ref{F8}(d)–(f). Consequently, the surface Chern numbers alone cannot fully characterize this redistribution of the chiral states.

To further reveal this mechanism, we analyze the evolution of the isolated surface bands in the $k_x$ momentum space. Figs.~\ref{F9}(a)-(d) show the isolated surface bands for $\phi=0$, $\phi=0.1\pi/17$, $\phi=0.5\pi/17$, and $\phi=0.8\pi/17$, respectively. In the weak-field regime [see Fig.~\ref{F9}(b)], the chiral modes can still be understood from the same surface-band-alignment mechanism as in the zero-field case [see Figs.~\ref{F4}(d)]. The weak magnetic field mainly modifies the separations between different surface bands in momentum space, without modifying their relative ordering [see Figs.~\ref{F9}(a) and (b)]. Based on the mechanism in Figs.~\ref{F4}(d), no qualitative transition occurs in the chiral edge states propagating along the $x$ direction for $\phi=0.1\pi/17$.

However, when the magnetic field reaches a certain strength, the Aharonov-Bohm phase changes the relative arrangement of the surface bands [see Figs.~\ref{F9} (c) and (d)]. After turning on the coupling between surfaces under the open boundary conditions, the induced chiral states undergo redistribution between different surfaces. Taking $\phi=0.1\pi/17$ and $\phi=0.5\pi/17$ as examples, the left/right surfaces ($\langle y \rangle = \pm N/2$) are shifted from both sides to the center. This surface band rearrangement drives the transfer of the transport channels from the $x$-$y$ surfaces to the $x$-$z$ surfaces in Figs.~\ref{F8}.

On this basis, we further quantify the relationship between the magnetic-field strength and the surface states in Figs.~\ref{F9}(a)-(d). After turning on the magnetic-field, the induced Aharonov-Bohm phases are summarized in Figs.~\ref{F9}(f). Notably, the front surface band is considered unshifted. In contrast, the back surface band is shifted in the $k_x$ momentum space by $\Delta k_x=-(L_z-1)\phi$. We denote the magnitude of this accumulated phase by $\phi_{\rm back}=(L_z-1)\phi$.  In the weak-field regime, the Dirac cones associated with the left and right surfaces lie approximately midway between those of the front and back surfaces.

To quantitatively capture the evolution of chiral states in real space, we plot the evolution of $\phi_{back}$ versus $\rho_{front}$. Here, $\rho_{front}=\sum_{k_x,y}|\psi(k_x,y,z=N/2)|^2$ counts the density of states located on the front surface, and $\psi(k_x,y,z=N/2)$ is the eigenstate of the chiral modes with $E = 0.067t$. As shown in Fig.~\ref{F9}(e), by increasing magnetic-field strengths, $\rho_{front}$ shows an obvious transition from almost zero to non-zero values. The transition occurs when the magnetic-field-induced relative momentum shift \(\phi_{\rm back}=(L_z-1)\phi\) becomes comparable to the zero-field separation \(d_x\) between the relevant surface Dirac cones [see Fig.~\ref{F9}(a) and the typical plot of $d_x$ is given in the inset of Fig.~\ref{F9}(e)]. This agreement clearly shows that the transition inherits its Chern-number independence from the chiral boudary states at zero magnetic field. Instead, the spatial evolution of the chiral modes in Figs.~\ref{F8}(a)-(c) originates from the competition between $\phi_{back}$ and $d_x$, which illuminates its intrinsic 3D nature.

\begin{figure*}[t]
  \centering
    \includegraphics[width=0.9\textwidth]{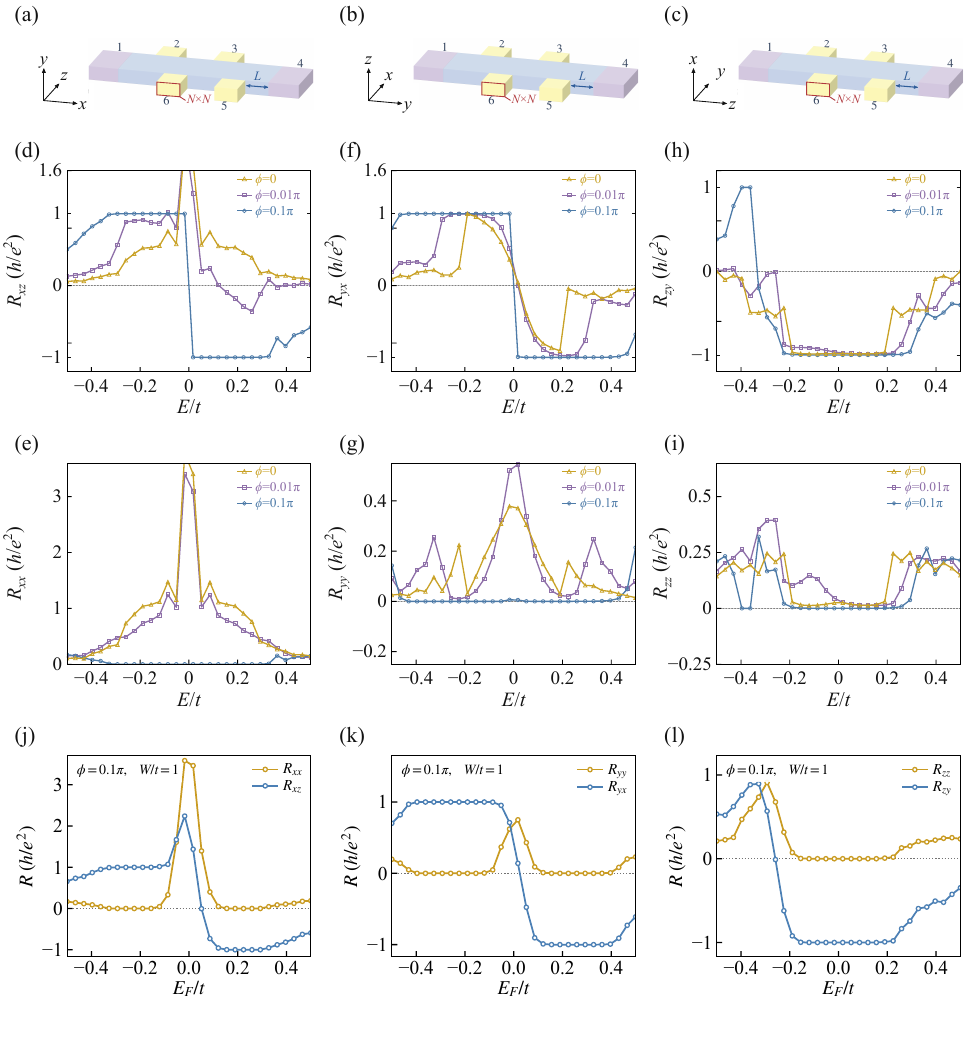}
    \caption{(a)--(c) Hall-bar setups. (d), (f), and (h) show the Hall resistances
$R_{\alpha\beta}=V_\beta/I_\alpha=(V_2-V_6)/I$, while (e), (g), and (i)
show the longitudinal resistances
$R_{\alpha\alpha}=V_\alpha/I_\alpha=(V_2-V_3)/I$, as functions of the
Fermi energy $E$. The corresponding current and magnetic-field directions
are $(x,y)$, $(y,z)$, and $(z,x)$, respectively. (j)--(l) show the
longitudinal and Hall resistances for the same three direction pairs,
respectively, at $\phi=0.1\pi$ and disorder strength $W/t=1$. The sample
dimensions are $N=10$ and $L=20$.}
    \label{F10}
\end{figure*}

Our results demonstrate that the transport properties along the $x$ direction respond directly to the relative positions of the uncoupled surface bands. This behavior agrees with the physical picture of the mechanism in Fig.~\ref{F4}(d), rather than the variation of Chern number. This observation further confirms that the new mechanism in Fig.~\ref{F4}(d) goes beyond the conventional Chern-number framework.

\section{Hall responses}\label{section5}

For clarity, we further present the simulations of Hall responses along all three directions. The typical Hall bar setups are shown in Figs.~\ref{F10}(a), (b), and (c) \cite{AsymmetricLeadNote}. The Hall responses are numerically calculated by adopting the multi-terminal methods in Sec. \ref{section2}(B) \cite{Buttiker1986FourTerminal,
Datta1995Electronic,PhysRevB.38.9375}.

Following the setup in Sec.\ref{section4}.C, we first focus on the case with a magnetic field along the $y$ direction. Here, the current is set along the $x$ direction. Figures~\ref{F10}(d) and (e) show the corresponding Hall resistance $R_{xz}$ and longitudinal resistance $R_{xx}$ as a function of the Fermi energy $E$ under different $\phi$. For $\phi=0$, $R_{xz}$ is nonquantized as $E$ increases, which is consistent with the absence of chiral modes on $x$-$y$ surfaces. For $\phi \neq 0$, $R_{xz}$ still deviates from the quantized features with $R_{xx}\neq0$ [see blue lines in Figs.~\ref{F10}(d) and (e)]. When $\phi$ is large enough, $R_{xz}$ captures the quantized values, and evolves from $h/e^2$ to $-h/e^2$ by increasing $E$. More importantly, when $R_{xz}$ exhibits a quantized plateau, the corresponding longitudinal resistance $R_{xx}=0$. These results demonstrate the existence of dissipationless chiral modes on the $x$-$y$ surfaces, only after the magnetic field is strong enough. Moreover, for quantized Hall responses, the Hall resistances are consistent with the Chern number and chiral mode distributions in Fig.~\ref{F8}.

Figures~\ref{F10}(f) and (h) show the Hall responses under magnetic fields oriented along the $z$ and $x$ directions, respectively. It clearly shows its 3D QAHE features, where the quantized $R_{yx}$ and $R_{zy}$ still hold for $\phi=0$. After turning on the magnetic fields, the evolutions of Hall resistances agree with the variation of Chern numbers and chiral states in Fig.~\ref{F5} and Fig.~\ref{F6}.

These quantized hall effects are robust against disorder, as shown in Figs.~\ref{F10}(j)--(l). In our calculation, the Anderson disorder $H_W=\sum_n\epsilon_n a^\dagger_na_n$ is adopted with $\epsilon_n$ in $\left[-W/2,W/2\right]$. $W$ is the disorder strength. Such robustness further unveils the topological nature of the three-dimensional quantum anomalous Hall effect under magnetic fields.

\section{Conclusion and discussion}\label{section6}

In summary, we have systematically investigated the magnetic-field-driven quantum Hall effect in a three-dimensional quantum anomalous Hall system with cross-dimensional transport, constructed by coupling two Weyl-semimetal Hamiltonians. Magnetic fields applied along different principal directions lead to highly anisotropic evolutions of the surface Chern numbers, chiral boundary modes, and Hall responses. For certain field and transport directions, the reconstruction of the chiral modes follows the evolution of the corresponding surface Chern numbers. In contrast, for the $x$-directed chiral modes under a $y$-directed magnetic field, the field modifies the relative momentum-space alignment of the surface bands and redistributes the chiral modes among different facets, even though the corresponding surface Chern numbers remain unchanged. This behavior reveals a breakdown of the conventional two-dimensional Chern-number-based bulk-boundary correspondence in this cross-dimensional geometry. The associated evolution of the Hall responses is further confirmed by multi-terminal transport calculations.

The underlying origin lies in the intrinsically three-dimensional coupling between different facets. Individual surfaces cannot in general be regarded as independent two-dimensional Chern insulators, because their surface states hybridize at the hinges and reconstruct the resulting chiral boundary modes. Consequently, the spatial distribution of the chiral modes is determined not only by the surface Chern numbers, but also by the relative momentum-space alignment and coupling of the surface bands. Our results clarify the interplay among surface topology, inter-facet coupling, and quantized Hall transport in three-dimensional cross-dimensional systems, and may motivate a more complete topological characterization of such states.

\section{acknowledgements}
We appreciate the valuable discussion with Ming Lu. This work was supported by the National Basic Research Program of China (Grants No. 2024YFA1409003 and No. 2022YFA1403700), National Natural Science Foundation of China (Grants No. 12350401).

\end{document}